# ACHIEVEMENTS AND NEW CHALLENGES FOR CERN'S DIGITAL LLRF FAMILY

M. E. Angoletta[†], S. Albright, A. Findlay, V. R. Myklebust, M. Jaussi, J. C. Molendijk, N. Pittet,
CERN, Geneva, Switzerland

*Abstract*

An innovative digital Low-Level RF (LLRF) family has been developed at CERN and deployed on several circular machines. Operation of CERN's PS Booster (PSB), Low Energy Ion Ring (LEIR) and Extra Low ENergy Antiproton (ELENA) ring all reaped great benefit from the flexibility and processing power of this new family. Beam and cavity feedback loops have been implemented, as well as bunch shaping, longitudinal blowup and bunch splitting. For ELENA, longitudinal diagnostics such as bunched beam intensity and bunch length measurements have also been deployed. During Long Shutdown 2 (LS2) the ferrite-based High-Level RF (HLRF) systems of the Antiproton Decelerator (AD) and of the four-ring PSB will be replaced with Finemet-based HLRF. This will require a new LLRF system for the AD and deep upgrades to the existing PSB LLRF systems. This paper gives an overview of the main results achieved by the digital LLRF family so far and of the challenges the LLRF team will take on during LS2.

## CERN VXS DIGITAL LLRF OVERVIEW

CERN's VXS Digital Low-Level RF (LLRF) family is the second generation digital LLRF for small synchrotrons developed at CERN. The first generation, now obsolete, operated successfully the Low Energy Ion Ring (LEIR) machine [1] and allowed carrying out machine tests in the Proton Synchrotron Booster PSB [2] for over 10 years.

The VXS digital LLRF family is based on the VXS bus and on CERN-designed custom hardware [3]. This includes: a) FMC-DSP carrier boards hosting up to two, high pin count FPGA Mezzanine Card (FMC) daughtercards; b) rear transition modules hosting the power supplies of the carrier board and providing timings and interlocks input-output capabilities; c) VXS switch modules allowing the various carrier boards in a crate to communicate; d) three types of FMC daughtercards providing the functions of Master Direct Digital Synthesizer (DDS), Slave DDS and Digital Down Converter. Figure 1 shows a FMC-DSP carrier board hosting two FMC daughtercards. The Field Programmable Gate Arrays (FPGAs) and the Digital Signal Processor (DSP) are also shown.

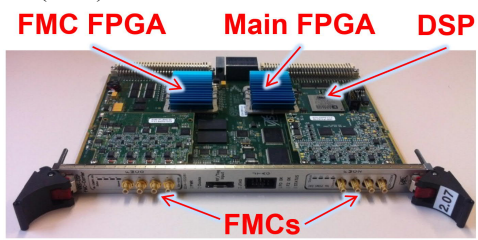

Figure 1: FMC-DSP carrier board.

The VXS LLRF family typically runs synchrotrons with low-frequency cavities (up to 20 MHz) hence it operates via direct sampling and baseband I,Q processing.

Initially based on a sweeping frequency clock, the LLRF is now being moved to fixed frequency operation [4]. This allows seamless handling of the frequency swing and improves the signal-to-noise ratio in the analogue-to-digital and in the digital-to-analogue conversion.

The family was successfully deployed on several CERN machines [5-7] and on the synchrotron of the MedAustron complex for hadron therapy [8]. It was also adopted by CERN's Beam Instrumentation group to implement orbit measurement systems [9, 10] and used for longitudinal diagnostics by CERN's Radiofrequency (RF) group [11]. Table 1 summarises the main deployment milestones, achieved and planned, together with the staff involved. The used clocking scheme (fixed vs. sweeping) is also indicated. Finally, studies to develop the 3$^{rd}$ generation LLRF family will start in 2023, after the post-Long Shutdown 2 (LS2) machines restart.

Table 1: VXS digital LLRF family deployment milestones. Keys: MA - MedAustron; RF - CERN BE Radiofrequency group; BI - CERN BE Beam Instrumentation group.

| When | What | Who |
|---|---|---|
| 2014 | MedAustron LLRF (sweeping clock) | MA, RF |
|  | PSB 4 rings LLRF (sweeping clock) | RF |
| 2016 | AD orbit | BI |
|  | LEIR LLRF upgrade to 2$^{nd}$ generation LLRF (sweeping clock) | RF |
| 2017 | ELENA orbit | BI |
|  | ELENA LLRF (fixed frequency clock) | RF |
| 2018 | LEIR orbit | BI |
|  | LEIR LLRF upgrade to fixed frequency clock | RF |
|  | ELENA LLRF upgrade to include some longitudinal diagnostics | RF |
| 2019 | **Long Shutdown 2** |  |
| 2020 | PSB LLRF upgrade to full Finemet HLRF operation and Linac4 injection (fixed frequency clock) | RF |
|  | ELENA LLRF upgrade to include full longitudinal diagnostics | RF |
| 2021 | AD LLRF and longitudinal diagnostics (fixed frequency clock) | RF |
| ~2022 | Beam loops implementation in PS LLRF (sweeping clock) | RF |
| ≥2023 | Start studies for 3$^{rd}$ generation LLRF | RF |

# PS BOOSTER

## The Machine

CERN's PSB is a Large Hadron Collider (LHC) injector, which accelerates protons with four superposed rings and supplies beam to an experimental area. It is being upgraded [12] as part of CERN's LS2 activities.

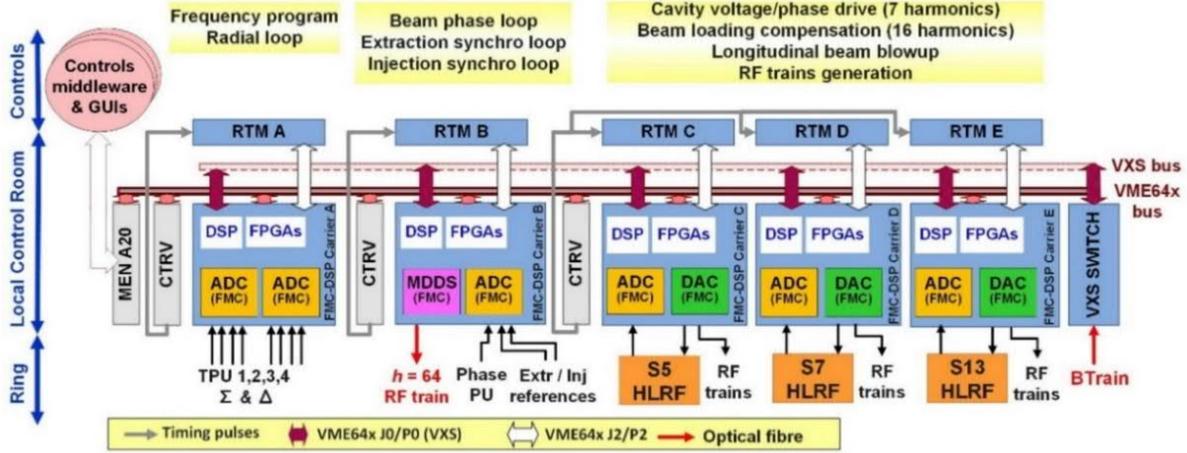

Figure 2: Post-LS2 LLRF for one PSB ring. Keys: MDDS – Master Direct Digital Synthesiser, ADC – Analogue-to-Digital Converter, DAC – Digital-to-Analogue Converter, TPU – Transverse Pick-Up, CTRV – timing receiver module, MEN A20 – Master VME board, RTM – Rear Transition Module.

## The LLRF

The PSB was successfully equipped in 2014 with one independent LLRF system per ring, based on the VXS LLRF family [5, 13]. These will be upgraded during LS2 to comply with new operational specifications. In particular, each LLRF will operate the three new Finemet-based HLRF systems per ring, that deliver up to 8 kV each in the (1 – 20) MHz bandwidth. Figure 2 shows the layout and functionalities of the LLRF system for one ring [14].

## Selected Achievements

Many new features were tested and reliability runs carried out over the years, particularly in view of post-LS2 operation [5, 13, 14]. Here two achievements are selected.

Table 2: Comparison of PSB longitudinal blowup methods: single tone modulation at high harmonic and phase noise.

| | | |
|---|---|---|
| High h | Pluses | Easy to track changing synchrotron frequency $f_S$ |
| | | Faster |
| | Minuses | Minimum of 5 D parameters space |
| | | Requires control of high harmonic |
| Phase noise | Pluses | No need for high h control |
| | | Smaller parameters space |
| | | Targets specific $f_S$ amplitudes |
| | Minuses | More difficult to track changing $f_S$. |
| | | Slower. |

A novel longitudinal blowup method by phase noise at the accelerating harmonic was validated in the PSB [15]. This method was so far used only in larger CERN machines such as the Super Proton Synchrotron (SPS) and the LHC.

On the contrary, the PSB operated with a single-tone modulation at high harmonic. Table 2 shows a comparison between the two methods, which will both be available after LS2.

Another achievement was operating with three harmonics ($h$ = 1+2+3) instead of two ($h$ = 1+2).

This was enabled by the test Finemet HLRF system installed in PSB Ring 4, that allowed operation also at harmonic $h$=3. Beam tests showed that triple harmonic operation improved the brightness of the LHC25 beam and reduced the vertical emittance of the Multi Turn Extraction (MTE) beam. Figure 3 depicts the simulated bunch profile for operation at triple harmonic, at standard double harmonic and at double harmonic with extra voltage on $h$=2 (double harmonic overloaded). The capability to control harmonic $h$=3 in voltage and phase is therefore a strong wish for the post-LS2 LLRF system.

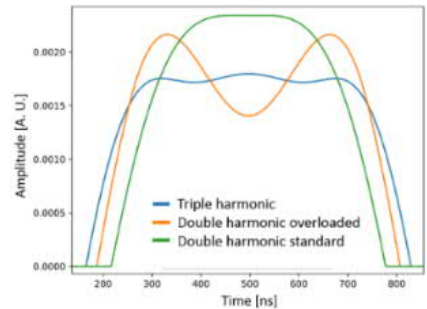

Figure 3: Simulated bunch profile for different harmonics used. Traces: triple harmonic [blue], double harmonic overloaded [orange], double harmonic standard [green].

## New Challenges

Post-LS2 challenges include operation with a new Btrain system, ring synchronisation for Linac4 injection and control of the Finemet-based HLRF systems. The last one is a major change as it includes implementing in the FPGA servoloops at 16 harmonics for each HLRF system. This approach has been the model for other accelerators [16].

# LEIR

## The Machine

CERN's LEIR is an LHC injector that accumulates and accelerates ions; particles accelerated so far include $O^{4+}$, $Ar^{11+}$, $Xe^{39+}$ and $Pb^{54+}$.

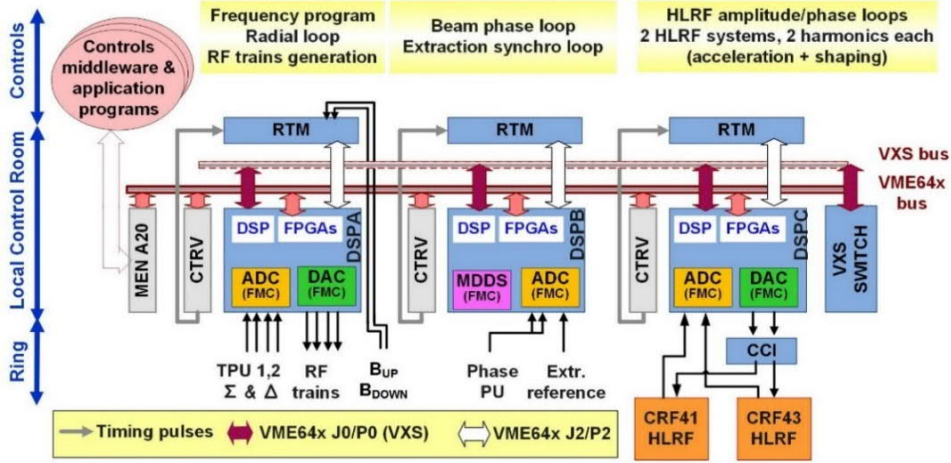

Figure 4: LEIR LLRF schematic view. Keys: MDDS – Master Direct Digital Synthesiser; ADC – Analogue-to-Digital Converter; DAC – Digital-to-Analogue Converter; TPU – Transverse Pick-Up; CTRV – timing receiver module; MEN A20 – Master VME board; RTM – Rear Transition Module; CCI – Cavity Control Interface.

## The LLRF

LEIR began commissioning in 2005 equipped with the 1st generation LLRF system [1]. It was upgraded to the 2nd LLRF generation with sweeping frequency clock operation in 2016 [7]. Finally, it was upgraded to fixed frequency clock operation in 2018. Figure 4 shows the LEIR LLRF layout and functionalities implemented. In particular, the LLRF operates in double-harmonic mode (acceleration and shaping) using either of the two HLRF systems installed in LEIR. Both HLRF systems can be also driven in parallel for machine development sessions. More details on the system operation are available elsewhere [7].

## Selected Achievements

The LEIR LLRF allowed many machine improvements [7]. Here three achievements are outlined.

First, a novel method for capturing ion beams [17] was routinely used for operation during the 2018 $Pb^{54+}$ run. This consists of a programmable modulation of the frequency during the capture, embedded in the LLRF.

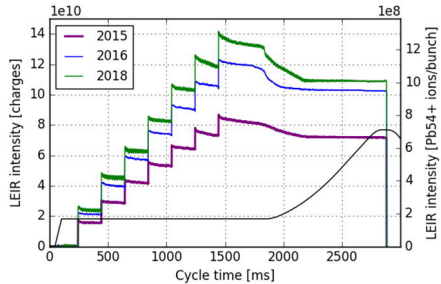

Figure 5: Progress in LEIR's extracted intensity for $Pb^{54+}$ ion operation from 2015 to 2018.

The method allowed improving both the reproducibility and the beam transmission through the machine. Figure 5 shows the progress in LEIR's extracted intensity in the last three $Pb^{54+}$ runs; the LEIR LLRF was instrumental in obtaining this result.

Second, a NOMINAL scheme operating at harmonics $h$=3+6 instead of $h$=2+4 was deployed in the 2018 run. With this novel scheme, originally unplanned, three bunches were successfully accelerated, synchronised, extracted (see Figure 6) and routinely sent through the ion accelerator chain to the LHC. This scheme will be the backup operational mode in case of problems with post-LS2 SPS momentum slip stacking operation.

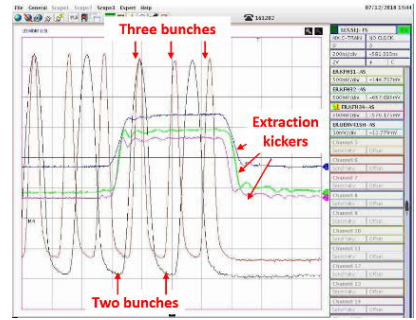

Figure 6: $Pb^{54+}$ NOMINAL bunches at extraction under the kickers. Traces: three bunches [brown], standard two bunches [black], extraction kickers [blue, green, violet].

Finally, tests were made with triple harmonic operation at $h$=2+4+6 and by using two cavities in parallel. Inspired by the PSB experience, this method showed a transmission slightly higher than with standard double harmonic. It will be further studied in future runs.

## New Challenges

Operation with the new Btrain system will be required after LS2. The automatic generation of LLRF parameters and voltage functions will be integrated within the LEIR controls infrastructure.

# ELENA

## The Machine

The Extra Low ENergy Antiproton (ELENA) ring [18] decelerates antiprotons injected from the Antiproton Decelerator (AD). For setting up it can also accelerate/decelerate H$^-$ ions and protons from a source. Its commissioning, still under way, began in December 2016. In 2018 ELENA started delivering antiprotons to GBAR [19].

## The LLRF

Figure 7 shows a schematic view of the ELENA LLRF layout and functionalities. In particular, the LLRF controls a Finemet-based HLRF capable of delivering up to 500 V. More details on the system are given elsewhere [6, 11].

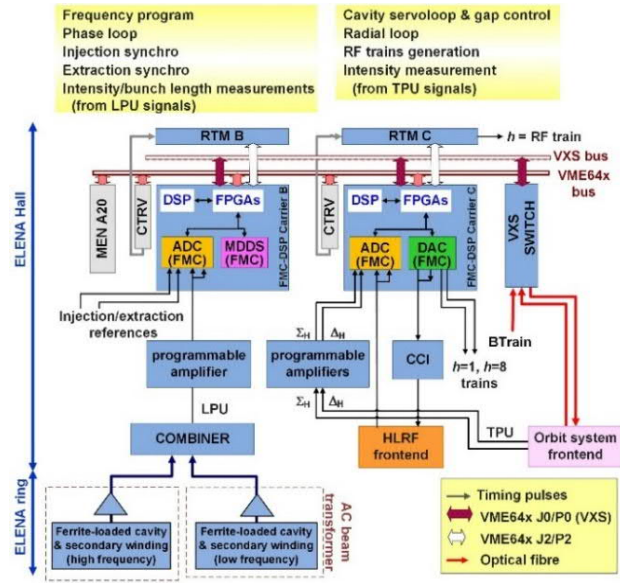

Figure 7: ELENA LLRF layout in 2018. Keys: MDDS – Master Direct Digital Synthesiser, ADC – Analogue-to-Digital Converter, DAC – Digital-to-Analogue Converter, CTRV – timing receiver module, MEN A20 – Master VME board, RTM – Rear Transition Module, CCI – Cavity Control Interface, LPU/TPU – Longitudinal/Transverse Pick-Up.

## Selected Achievements

ELENA's LLRF allowed to carry out bunch-to-bucket transfer from the AD, to decelerate antiprotons as well as H$^-$ ions and to synchronise at extraction the bunch(es). [6, 11]. Here previously unpublished results are shown.

The H$^-$ beam was routinely accelerated and decelerated to setup the machine. Figure 8 shows the cycle and several LLRF signals such as the magnetic field, the phase and radial loop contributions in units of Hz, the radial loop position in units of mm measured by the LLRF and received by the orbit system over optical fibre.

Figure 9 shows the bunch length measurement for an antiproton cycle as an example of the longitudinal diagnostics developed for ELENA. As expected, the bunch length increases during the acceleration and decreases on the extraction plateau, also thanks to the bunched-beam cooling. The discontinuity in the bunch length on the first ramp is due to a phase jump on the signal changeover between high-frequency and low-frequency LPU.

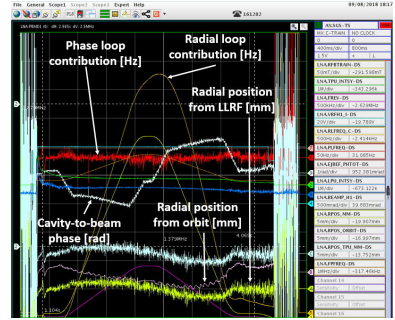

Figure 8: Accelerated and decelerated H$^-$ beam.

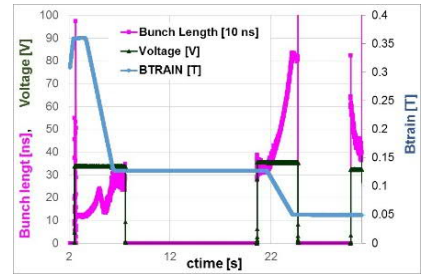

Figure 9: Antiprotons bunch length measurement. Traces: bunch length [pink], magnetic field [blue], voltage [green].

## New Challenges

The longitudinal diagnostics will be expanded with the ObsBox, a custom processing module based on a server PC, with high data rate optical interfaces and large storage capabilities [20]. The resulting diagnostics will be exported first to the AD, then to other machines. The combination of signals in the LPU will be corrected as a function of frequency. Finally, the control of the LLRF parameters will be integrated within the RF cycle editor, to automatically adapt the parameters to different cycles.

# AD

## The Machine

CERN's AD has been providing antiprotons to experiments since July 2000. It is now being upgraded and consolidated [21].

## The LLRF

The AD will restart after LS2 equipped with a customised copy of ELENA's LLRF as its new digital LLRF system. This will provide beam and cavity control as well as longitudinal diagnostics. Figure 10 shows the system layout and its functionalities. In particular, the LLRF will control a Finemet-based HLRF delivering up to 3500 V [20].

## Challenges

The AD will profit from ELENA's LLRF features and commissioning experience. The AD commissioning will however have to compete for manpower with the PSB commissioning and the restart of other LHC injectors.

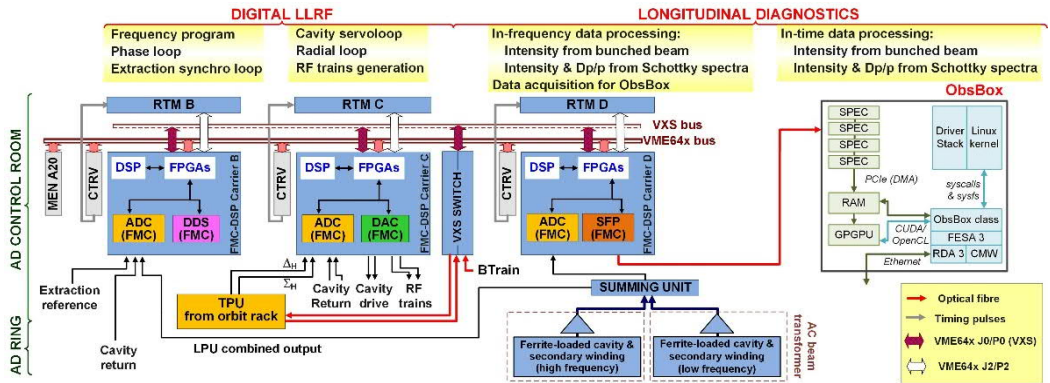

Figure 10: Post-LS2 AD LLRF and longitudinal diagnostics. Keys: DDS –Direct Digital Synthesiser, ADC – Analogue-to-Digital Converter, DAC – Digital-to-Analogue Converter, SFP – Small Form-factor Pluggable Transceiver, LPU/TPU – Longitudinal/Transverse Pick-Up, CTRV – timing receiver module, MEN A20 – Master VME board, RTM – Rear Transition Module, ObsBox – custom processing module, SPEC – Simple PCI Express Carrier module.

## CONCLUSION AND FUTURE WORK

Developing CERN's VXS LLRF family was a large RF group manpower investment, well repaid by synergies amongst the various machines this LLRF operates. Its flexibility and processing power has allowed to implement features not originally planned and to fully harness the wideband characteristics of the Finemet HLRF. The RF group successfully used this family on ELENA for longitudinal diagnostics, which will be expanded and used by other machines. Finally, studies for the 3rd generation LLRF family will start in 2023, after the post-LS2 machines restart.

## ACKNOWLEDGEMENTS

We are grateful to our colleagues in the operation and beam physics groups for their support. The fruitful collaboration with HLRF colleagues is also acknowledged.